# Study of point spread in aberration-corrected high-resolution transmission electron microscopy

B. H. Ge[*], Y. M. Wang, Y. Yao, F. H. Li

*Beijing National Laboratory for Condensed Matter Physics,*
*Institute of Physics, Chinese Academy of Sciences, Beijing 100190*

## Abstract

For quantitative electron microscopy high precision position information is necessary so that besides an adequate resolution and sufficiently strong contrast of atoms, small width of peaks which represent atoms in structural images is needed. Size of peak is determined by point spread (PS) of instruments as well as that of atoms when point resolution reach the subangstrom scale and thus PS of instruments is comparable with that of atoms. In this article, relationship between PS with atomic numbers, sample thickness, and spherical aberration coefficients will be studied in both negative $C_s$ imaging (NCSI) and positive $C_s$ imaging (PCSI) modes by means of dynamical image simulation. Through comparing the peak width with different thickness and different values of spherical aberration, NCSI mode is found to be superior to PCSI considering smaller peak width in the structural image.

[*] Corresponding author. Email: bhge@iphy.ac.cn

# 1. Introduction

When aberrations of objective lens in electron microscopes are corrected, and then resolution is sufficient to solve all atoms, precision becomes mandatory for quantitative high-resolution electron microscopy (HREM). According to the study of Van Aert et al [1], precision of atomic position obtained by means of HREM depends on atomic distance, the resolution of microscope, dose of electron counts and width of peaks which represent atoms.

For a conventional transmission electron microscope usually with the point resolution about 2 Å, peak width in Scherzer-focus images (structural images) is mainly limited by the point spread (PS) of instruments, the Fourier transformation of phase contrast transfer function of objective lens [2], according to the weak-phase object approximation, in which image intensity is convolution of projected potential of samples with PS of instruments. But for an aberration-corrected microscope, point resolution is reduced to even 0.5 Å, and PS caused by instruments is comparable with that caused by objects so that contribution of the latter cannot be ignored in determining the peak width [3]. Simulation confirms the infinite point spread introduced by projected potential of objects [3], termed as object point spread, which will limit the width of peak in the image and also the ultimate 'structure resolution'.

In addition, in aberration-corrected electron microscopy, besides the positive $C_s$ imaging (PCSI) mode tunability of the spherical aberration makes available a new imaging mode, negative $C_s$ imaging (NCSI) mode, negative value of spherical aberration together with an overfocus [4-6]. Peak width in NCSI mode has been found to be obviously smaller than that in PCSI one in observation of the oxide superconductor $Bi_2Sr_{1.6}La_{0.4}CuO_{6+\delta}$, which has been verified by image simulation [7].

In this article, influencing factors as to the PS in transmission electron microscopy is discussed in detail including atomic number, sample thickness, and spherical aberration coefficients as well as two imaging modes and is studied not only in image plane but also in projected potential and exit plane.

For the difficulty of studying the PS analytically, the numerical method, image simulation comprising a full dynamical calculation of electron scattering and the contrast transfer under partially coherent illumination, is utilized, and width of peaks in images is used to describe the PS.

# 2. Image simulation

As a general case, fictitious model with face-centered cubic structure (atomic position (0, 0, 0), (0.5, 0.5, 0)) is used to investigate the PS in this article. To avoid the interaction between adjacent atoms lattice parameter of the model is set to be $a$ =0.41 nm and then the smallest atomic distance is about 0.2 nm. Atomic species was varied to investigate the effect of scattering power, and a Debye-Waller factor of 0.005 nm$^2$ has been applied regardless of the atomic species in the simulation of the Multislice method. The main imaging parameters were an accelerating voltage 300 kV, energy spread 0.8 eV, semiconvergence angle 0.1 mrad and vibration 0.03 nm with no other aberrations considered.

In addition, when values of the spherical aberration are corrected to several tens of microns or less, the Scherzer focus value [10] is comparable with the sample thickness. In this case, the value should be corrected using an effective object plane close to the midplane instead of the exit plane of sample [3, 8, 9], that is $\Delta f_{mid} = \Delta f_{ex} + 0.5t$, where $\Delta f_{mid}$ and $\Delta f_{ex}$ indicate the focus value at the midplane of the sample and the exit plane, respectively, and $t$ is sample thickness. For example, for $Cs = -13$ μm, the Scherzer focus is calculated to be 6.2nm according to the results of Scherzer [10], but when sample thickness, say 6 nm, is taken into consideration, the corrected Scherzer focus should be 3.2 nm including the underfocus $-t/2$.

Moreover, according to the pseudo-weak-phase object approximation [11], with increases of sample thickness intensity of peaks in structural images may deviate from accumulated atomic number along the electron beam direction, but the position of peaks is accurately same as the position of atoms in both NCSI and PCSI modes. In this article, the intensity of peaks is not considered such that simulated images with the Scherzer focus condition can be regarded as structural images.

## 3. Results

### 3.1. Peak width in projected potential maps (PPMs) with different atomic number

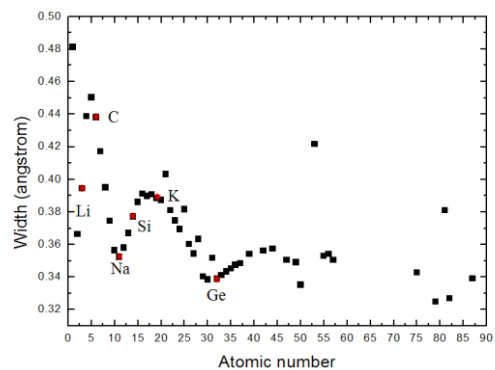

Fig. 1 Peak width in projected potential maps versus atomic number.

Peak width in potential maps is determined by the Lorentz function fitting, and the scatter plot of peak width versus atomic number is shown in Figure 1. Oscillation of peak width with no obvious period was found with increasing atomic number, as well as no obvious relationship between the atomic scattering power [12] and peak width, which is different with results in Ref. [3]. This chaos may be originated from the complex correlation between the scattering power and the potential. That is, scattering power, a variable in the reciprocal space, is a main parameter of atomic structure factor, while potential, a variable in the real space, is the Fourier transformation of the structure factor.

## 3.2 Peak width in the exit plane with different thickness.

When electrons transmit through materials, especially crystals in the main axis, their trajectories will be changed due to the electrostatic potential of the screened nuclear. Thus, the distance for electrons to transmit in the materials, i.e. the sample thickness, will influence the peak width in the exit plane. Six atomic species Li, C, Na, Si, K and Ge with different scattering power were selected in random as marked in Fig. 1 to study the PS in the exit plane.

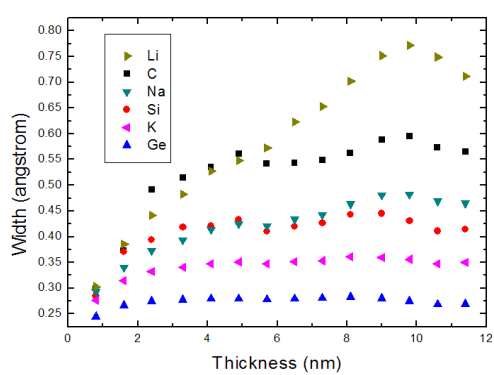

Fig. 2 Peak width in the intensity of exit waves versus sample thickness for Li, C, Na, Si, K and Ge, respectively.

Peak width in the intensity of exit waves is determined by the Gaussian function fitting, and variations of peak widths for six atom species with different thickness are shown in Fig. 2. Although there is some disorder when thickness is smaller than 6 nm, it is clear that with more thickness atoms with larger atomic number corresponds to smaller peak width, which is different from the results in Fig. 1. Moreover, the result in Fig. 2 is also contrary to our perspective that heavy atoms should be with large peak width. In the following subsection it will be shown that this perspective is usually right for PCSI mode but not for NCSI one. In addition, in Fig. 2 it can be observed that with increase of sample thickness there is some fluctuation for peak width, but no uniform period was found after detailed study with smaller thickness interval and with much more thickness.

Otherwise, considering the increasing intensity of peak with increasing thickness when sample is very thin [11] in combination with the increasing width as shown in Fig. 2, the number of electrons taking part in imaging seems to increase with increase of thickness, which is paradoxical. In reality, however, with increasing thickness, more electrons are attracted to take part in diffraction such that transmission beam is weakened in the reciprocal space and then background in the image will be reduced. Therefore, it is the effective electron (not forming the background) number that increases.

## 3.3 Peak width in the image plane with different thickness

Electrons are reflected when through objective lenses to form images such that peak width in the image plane will be influenced by imaging conditions such as the spherical aberration. In this and next subsections sign and magnitude of the spherical aberration combined with the sample thickness will be taken into consideration, respectively, and C, Si and Ge are selected as examples. Gaussian function fitting is used in these two subsections to determine the peak width.

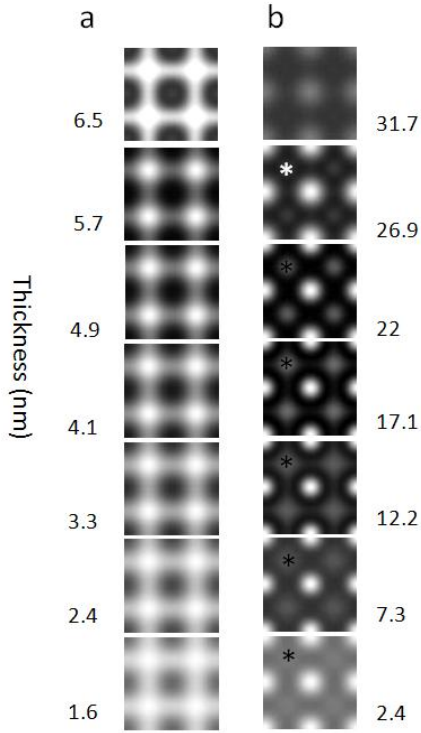

Fig. 3 (a) and (b) Thickness series of simulated image of Si with $C$s = 0.013 and -0.013 mm, respectively. Fictitious spots are denoted by stars. To show the contrast reverse in the image of (a) with thickness 6.5 nm, brightness is increased.

Thickness series of images for Si are shown in Figs 3(a) and (b), PCSI mode with $C$s = 0.013 mm and NCSI mode with $C$s = -0.013 mm, respectively. Scherzer focus condition is corrected as mentioned in section 2. In Fig. 3(a) black contrast can be interpreted as atoms, while in Fig. 3(b) white contrast. Obvious discrepancies can be observed in Figs. 3(a) and (b). Firstly, the critical thickness [11], which means with that thickness the contrast of dots corresponding to atoms is close to zero, is about 6 nm in PCSI mode (see Fig. 3(a) (to show the contrast reverse in the image with thickness 6.5 nm, brightness is increased), but for NCSI (see Fig. 3(b)) the critical thickness is about 32 nm. Secondly, in PCSI mode, peak width monotonously increases with increasing thickness below the critical thickness while in NCSI mode the peak width decreases at first with increase of thickness and then increases. Similar results can be obtained for C and Ge, which can be seen clearly in Figs. 4(a) and (b), the scatter plot of peak width vs. thickness in PCSI and NCSI, respectively. Moreover, from Fig. 4 one feature can be found that the peak width of the atom with the larger atomic number increases more rapidly in PCSI but in NCSI it decreases more rapidly when sample is not very thick, say 10 nm. Thus, the larger atomic number, the larger peak width in PCSI, while in NCSI it is opposite.

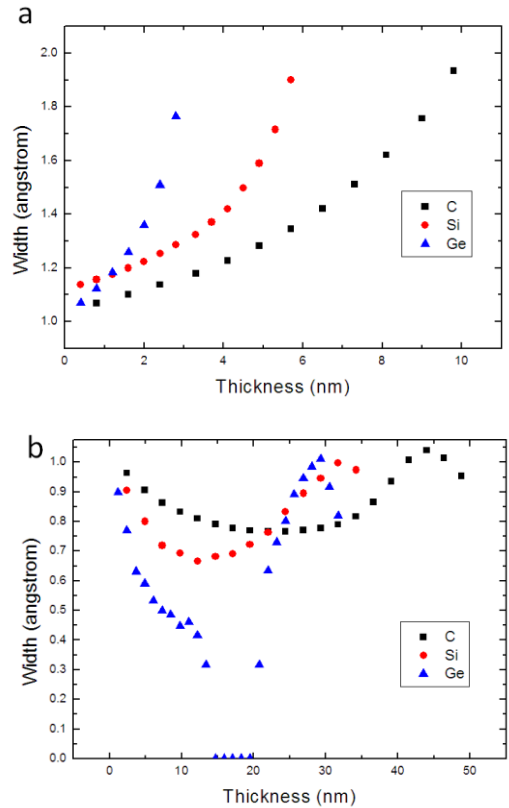

Fig. 4 (a) and (b) Scatter plot of peak width for C, Si and Ge in simulated image versus sample thickness with $C$s= 0.013 and $C$s = -0.013 mm, respectively。

Things become more complicated for NCSI when sample is too thick that peak width fluctuates with thickness increasing, which is similar as shown in the exit plane, but the maximum peak width in NCSI is always smaller than the minimum width in PCSI at least in the scope of our research. From this point of view, NCSI is superior to PCSI.

But on the other hand, fictitious spots in the NCSI mode can be found as shown in Fig. 3(b), one of which in each image is denoted by a star, while in the mode of PCSI as shown in Fig. 3(a) it is not the case. Contrast of the fictitious spots is comparable with that of true structure at some thickness, which may make the image not directly interpretable. It will be shown in the next subsection that with increasing magnitude of the spherical aberration coefficient contrast of fictitious spots in the NCSI mode can be weakened.

### 3.4 Peak width in the image plane with different $C_s$ values

Figs. 5 (a) and (b) shows $C_s$ series of images of Si with sample thickness 2.4 nm in PCSI and NCSI modes, respectively. In general, the peak width decreases with decreasing magnitude of $C_s$ values for both PCSI and NCSI modes. More details can be seen clearly in Fig. 6, the scatter plots of peak width versus $C_s$ with three thickness value 1.6, 2.4 and 3.3 nm. In Fig. 6(a), the PCSI mode, the rapid decrease of the peak width is found when magnitude of $C_s$ decreases from 0.1 mm to 0.05 mm, but with further decreasing no obvious decrease (even increase). On the contrary, in NCSI mode (see Fig. 6(b)) peak width decreases gradually with the decreasing of the magnitude of $C_s$ from 0.1 mm to 0.05 mm and then decreases steeply from 0.013 to 0.003 mm. Moreover, from Fig. 6, it can be seen that peak width in NCSI mode is always smaller than in PCSI mode with any $C_s$ value. Combining the results from Fig. 6 and those from Fig.5, to obtain smaller peak width and better structural resolution NCSI is preferred.

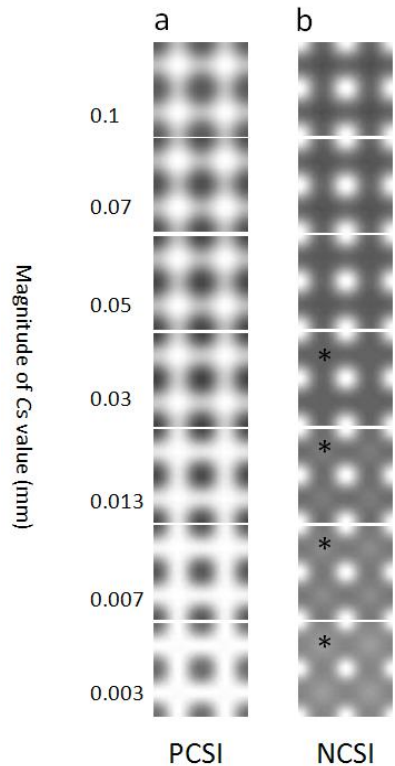

Fig. 5 (a) and (b) $C_s$ series of simulated image of Si with thickness 2.4 nm in both the PCSI and NCSI modes, respectively. Fictitious features are denoted by stars.

Nevertheless, similarly as mentioned in the last subsection, there appear fictitious spots in the image of NCSI mode as shown in Fig. 5(b) when $C_s$ is close to zero, and with its decreasing contrast of fiction become more and more obvious, vice versa. Hence, more attention should be paid to interpret the image contrast in NCSI mode.

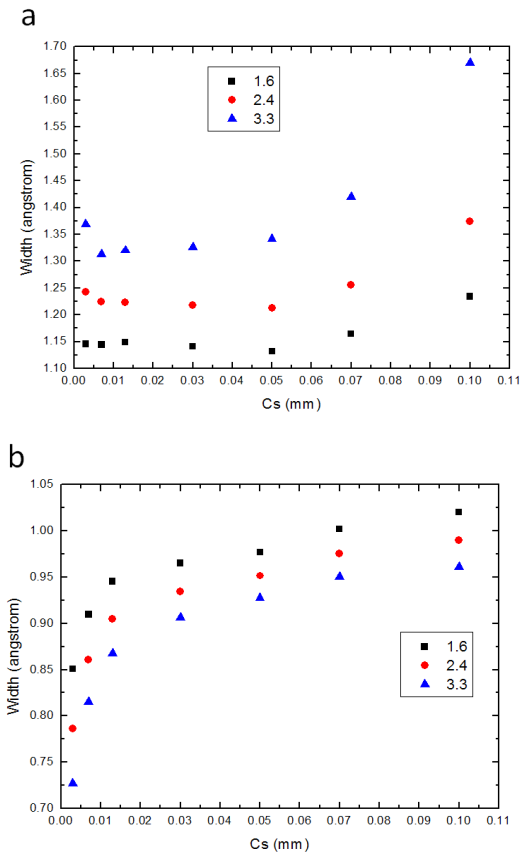

Fig. 6 (a) and (b) Scatter plots of peak width with three thickness values (1.6, 2.4 and 3.3 nm) versus $C$s values in PCSI and NCSI modes, respectively. Sign of $C$s for NCSI is neglected.

## 4. Discussion

NCSI is newly developed imaging mode in the aberration-corrected electron microscopy, and has shown incomparable superiority over PCSI as follows. First, through choosing optimum imaging condition, Scherzer defocus condition and lichte defocus condition [13] can be satisfied at the same time[14] such that a compromise between a reduced delocalization and a high mount of phase contrast can be obtained. Second, due to the constructive superposition of linear and non-linear contribution to the total image contrast [5, 15], the image contrast has been greatly improved in NCSI mode as well as signal against noise caused by amorphous surface layers [16]. Thus it has been wide used to study weakly scattering atoms surrounded by strongly scattering atoms [4-6]. Third, critical thickness [11], also named the extinction distance, is thicker in NCSI than that in PCSI if only the sign of spherical aberration is different as shown in Fig. 3. Thus, with more sample thickness the image can be directly interpreted in NCSI than PCSI. Moreover, the contrast reversal in NCSI mode do not appear like in PCSI, which will make the image contrast in NCSI mode more interpretable even when the thickness exceeds the critical thickness. Another advantage for NCSI is mainly discussed in this article, the smaller peak width in structural images. Especially, with the increase of thickness the peak width decreases first in NCSI, while it always increases in PCSI.

Although the NCSI mode is preferred in structural determination by HREM, attention still should be paid to interpret the image contrast in NCSI mode, for fictitious feature may appear when $C$s is close to zero as shown in Figs. 3(b) and 5(b). Fake feature can be corrected by posterior image processing [17]. Alternatively, larger $C$s value can be chosen to avoid it.

## 5. Summary

Point spread described with the peak width in this article has been investigated by dynamical image simulation in different stage, PPM, exiting plane and imaging plane. Following conclusions have been reached:

1、 No obvious relationship was found

between peak width in PPM with atomic number as well as scattering power.
2、Smaller peak width for heavy atoms than for light ones has been found in exit wave plane, which is different with the results in PPM.
3、With increasing thickness peak width in the image plane decreases first in the NCSI mode, but it always increases for the PCSI one. And with increasing thickness width of peak with larger atomic number increases more rapidly in PCSI, while in NCSI it decreases more rapidly.
4、With decreasing magnitude of $C_s$ values peak width decreases generally in both NCSI and PCSI modes.

In general, with the same condition peak width in NCSI mode is always smaller than that in PCSI mode. Considering peak width is one of dependent factor of atomic position precision, NCSI is preferred for quantitive structural determination. Due to the appearance of fictitious spots in NCSI mode when $C_s$ is close to zero, however, one can choose a reasonable larger $C_s$ value with the sacrifice of resolution.

## Acknowledgements

This work was supported by the National Basic Research Program of China (grant number: 2011CBA001001) and the National Natural Science Foundation of China (grant number: 50672124, 10874207 and 11104327).

[1] S. Van Aert, A.J. den Dekker, D. Van Dyck, A. van den Bos,  J. Struct. Biol., 138 (2002) 21.
[2] E.J. Kirkland, Advanced Computing in Electron Microscopy, Springer, 2010.
[3] M. Lentzen,  Microsc. Microanal., 14 (2008) 16.
[4] C.L. Jia, M. Lentzen, K. Urban,  Science, 299 (2003) 870.
[5] C.L. Jia, M. Lentzen, K. Urban,  Microsc. Microanal., 10 (2004) 174.
[6] C.L. Jia, K. Urban,  Science, 303 (2004) 2001.
[7] B. Ge, Y. Wang, F. Li, H. Luo, H. Wen, R. Yu, Z. Cheng, J. Zhu,  in preparation.
[8] P. Bonhomme, A. Beorchia,  J. Phys. D: Appl. Phys., 16 (1983) 705.
[9] R. Yu, M. Lentzen, J. Zhu,  Ultramicroscopy, (2011).
[10] O. Scherzer,  J. Appl. Phys., 20 (1949) 20.
[11] F.H. Li, D. Tang,  Acta Cryst. A, 41 (1985) 376.
[12] P.A. Doyle, P.S. Turner,  Acta Crystallogr. Sect. A, 24 (1968) 390.
[13] H. Lichte,  Ultramicroscopy, 38 (1991) 13.
[14] M. Lentzen, B. Jahnen, C.L. Jia, A. Thust, K. Tillmann, K. Urban,  Ultramicroscopy, 92 (2002) 233.
[15] M. Lentzen,  Ultramicroscopy, 99 (2004) 211.
[16] C.L. Jia, L. Houben, A. Thust, J. Barthel,  Ultramicroscopy, 110 (2010) 500.
[17] M. Texier, J. Thibault-Pénisson,  Micron, 43 (2012) 516.